\documentclass{llncs}
\usepackage{llncsdoc}

\usepackage[english]{babel}
\usepackage{latexsym}
\newcommand{\cancel}[1]{}
\usepackage[caption=false]{subfig}

\begin{document}
\title{An All-Or-Nothing Flavor to\\ the Church-Turing Hypothesis}
\author{Stefan Wolf}
\institute{Faculty of Informatics, Universit\`{a} della Svizzera
  italiana (USI), CH-6900 Lugano.\\
 Facolt\`a indipendente di Gandria, Lunga scala, CH-6978 Gandria.}

\maketitle

\begin{abstract}
	\noindent
 Landauer's 
principle claims that ``Information is Physical.''
It is not surprising that
its conceptual {\em antithesis\/}, Wheeler's ``It from Bit,'' 
has  been more popular among computer scientists~| in the 
form of the {\em Church-Turing  hypothesis:\/}
 All natural
processes
can be computed by a universal Turing machine;  physical laws then
become  descriptions of subsets of {\em observable}, as opposed to merely
{\em possible}, computations. 
Switching back and forth between the two traditional styles of thought,  
motivated by quantum-physical Bell correlations
and the  doubts they raise about fundamental  space-time causality, we look for an
intrinsic, physical 
randomness notion and find one around the second 
law of thermodynamics.
Bell correlations 
combined with {\em complexity as randomness\/} tell us
 that beyond-Turing computations are
either
physically impossible, or they can be carried out by ``devices'' as
simple as individual  photons.
\end{abstract}

\section{Introduction}

\subsection{Ice {\em versus\/} Fire}
According to {\em Jeanne Hersch\/}~\cite{jh}, the entire history of philosophy 
is coined by an   antagonism rooting in the fundamentally 
opposite world views of the pre-Socratic philosophers 
{\em Parmenides of Elea\/} (515 B.C.E. -- 445 B.C.E.)
on the one hand and {\em Heraclitus\/} (520 B.C.E. -- 460 B.C.E.)
on the other.
For Parmenides, any change, even time itself, is simply an {\em illusion}, whilst for Heraclitus, 
{\em change\/} is all there is. The ``cold logician'' Parmenides has been compared
to {\em ice}, and Heraclitus' thinking is the {\em fiery\/}
one~\cite{nitsch}. 
If Hersch is right, and this opposition between these
styles
crosses the history of philosophy like a red line, then 
this must be true no less for the history of {\em science}. 

A textbook example illustrating the described antagonism is the debate between 
{\em Newton\/} and {\em Leibniz\/}~\cite{cl}: 
For Newton, space and time are fundamental and given {\em a
  priori}, just like a stage on which all the play  is
located. For Leibniz, on  the other hand, space and time are emergent
as {\em relational\/} 
properties: The stage emerges {\em with\/} the play and not prior to it, and it is not there
without it. 
With only a few exceptions~| most notably {\em Ernst Mach\/}~| 
the course of physical  science went for Newton's view; it did so with good
success. 
An important example here is, of course, Einstein's
relativity: Whilst its crystallization point was {\em Mach's
  principle}, 
stating that intertial forces are relational (as opposed to coming
from 
acceleration against an {\em absolute\/} space), the resulting theory does 
{\em not\/} follow the principle since there is (the flat) space-time 
also   in a massless universe. 

In the present work, we turn our attention to physical phenomena such as 
the second law of thermodynamics and Bell correlations from quantum
theory. 
We find here again the opposition between
Parmenides' and Heraclitus' standpoints, and  
we directly  build on their tension  with the goal of obtaining more insight, hereby
bridging the gap separating them to some extent. 

 The {\em Heraclitean\/} style can be recognized again in the spirit of {\em Ferdinand
  Gonseth\/}'s ``La logique est tout d'abord une science naturelle''~|
``Logic
is, first of all, a natural science.''  This is a predecessor of
Rolf Landauer's  famous slogan ``Information Is Physical,''~\cite{landauer61}
putting physics at the basis of the concept of information and 
its 
treatment.  

This is in sharp contrast to {\em Shannon\/}'s~\cite{shannon49} (very
successful) making information  {\em abstract}, independent of the 
particular physical realization of it ({\em e.g.}, a specific noisy communication
channel). 
To the {\em Parmenidean\/} paradigm belongs also the {\em Church-Turing
  hypothesis\/}~\cite{CT}, stating that all physically possible processes can be
simulated by a universal Turing machine. This basing physical reality
on information and computation was later summarized by {\em John Archibald
  Wheeler\/} as ``It from Bit''~\cite{wheeler}.

\subsection{Non-Locality, Space-Time Causality, and Randomness}

After Einstein had made the world mechanistic and ``local'' (without
actions at a distance), 
he was himself involved in a work~\cite{EPR} paving the (long) way to
that locality to fall again. The goal of Einstein {\em et al.}\ had,
however,
been the exact opposite: to {\em save\/} locality in view of correlations 
displayed in the measurement behavior of (potentially physically
separated) parts of an 
{\em entangled\/} quantum state. 
The claim was that quantum theory was an only incomplete
description of nature, to be refined by
hidden parameters determining 
 the outcomes of all 
potential, alternative measurements. It took roughly thirty years
until that claim was grounded when {\em John Stewart Bell\/}~\cite{bell} showed
the
impossibility of the program~| ironically making the case with the
exact same states
as ``EPR'' had introduced. The consequences of Bell's insight
are  radical: If the values are not predetermined, then there 
must be {\em fresh\/} and at the same time {\em identical\/} pieces of 
classical information popping us spontaneously~| this is {\em
  non-locality}.
The conceptual problem  these correlations lead us into is 
the difficulty of
explaining their origin {\em causally}, {\em
  i.e.},
according to {\em Reichenbach's principle\/}~| which states  that a
correlation between two space-time events
can stem from a {\em common cause\/} (in the common past) or a {\em direct influence\/}
from one event to the other~\cite{Reichenbach}. 
Bell's result rules out the common cause as an explanation, thus remains 
the influence. 
Besides the fact that it is an inelegant overkill to explain a {\em non\/}-signaling
phenomenon 
({\em not\/} allowing for transmitting messages from one party to the
other) using a {\em signaling\/} mechanism,
there are  further problems: Hidden influences as explanations
of Bell correlation require both infinite speed~\cite{beforebefore},~\cite{coretti},~\cite{tomy} and
precision~\cite{ws}. 

In view of this, it appears reasonable to question the (only)
assumption made in Reichenbach's principle: The  {\em a priori\/} 
causal structure~\cite{ocb}.\footnote{It has been shown~\cite{aemin}
  that if causality is dropped
 but logical
  consistency maintained, then a rich world opens~| comparable to the 
one between locality and signaling.}
 If we turn back the wheel to the
{\em Newton-Leibniz debate}, and choose to
follow 
Leibniz instead, seeing space-time as appearing only {\em a
  posteriori}, then there is a first victim to this: {\em Randomness\/}: 
In~\cite{colrenamp}, a piece of information is called 
{\em freely random\/} if it is statistically independent from all 
other pieces of information except the ones in its {\em future\/} light cone.
Clearly, when the assumption of an initially given causal structure 
is dropped, such a definition is not founded any
longer.\footnote{Note 
furthermore that the definition  is
consistent with full determinism: 
A random variable with trivial distribution is independent of every
other
(even itself).}
(It is then
possible to turn around the affair and base past and future on
postulated freeness of bits~\cite{njp}.)
In any case, we are now  motivated to find an
{\em intrinsic,
context-free, physical definition of randomness\/}  and choose to
look~at: {\em Watt's Steam Engine}.

\section{The Search for an Intrinsic Randomness Notion: From Steam Pipes to the 
Second Law of Thermodynamics}

\subsection{The Fragility and the Robustness of the Second Law}

The {\em second law of thermodynamics\/} has advanced to
becoming pop
culture.\footnote{See, {\em e.g.}, Allen, W.,  Husbands and Wives
  (1992): The protagonist
  Sally is explaining why her marriage did not work out. First she
  does not know, then she realizes: ``It's the {\em second law of
  thermodynamics\/}: sooner or later everything turns to shit. That's my
  phrasing, not the {\em Encyclopedia Britannica}.''} 
It is, however, much less famous than Einstein's relativity,
Heisenberg's uncertainty, or quantum teleportation because it does not have 
any glamour, fascination, or hope attached to it: The law stands for
facts many of us are  in  denial of or try to escape. 
We ask whether the
attribution of that formalized pessimism to physics
has {\em historical\/} reasons.

The validity of the second law  seems to depend
on 
surprising conditions such as
the inexistence of certain life styles ({\em e.g.},  {\em Maxwell's
demon\/}
or 
photosynthesizing
plants~| Kelvin~\cite{kelvin} writes: ``When light is absorbed {\em other 
than in vegetation}, there is dissipation [...]'').
To make things worse, there is always a non-zero probability  (exponentially
small, though) of
exceptions
where the law fails to hold; we  are not used to this from other
laws of physics. Can this be taken as an indication that the fundamental way of
formulating the law  eludes us?

The described {\em fragility\/} of the second law is strangely
contrasted by
its being, in another way, {\em more robust\/} than others (such as Bell 
violations only realizable under extremely precise lab conditions): We certainly do not need 
to trust experimentalists to be convinced
that the second law is acting, everywhere and always. It has even been
claimed~\cite{uffink} to hold a ``supreme position''  among
physical 
laws: It appears easier to imagine a world where relativity or quantum
theory 
do not hold than to figure out a reality
lacking the validity of the second law. (Concerning the reasons for
this, we can only speculate: Would we be forced to give up the mere {\em
  possibility of perception,  memory~| the arrow of time\/}?)

\subsection{History}

This story (see~\cite{uffink}) starts with {\em Sadi Carnot\/}
(1796--1832) and
his
study of heat engines such as {\em James Watt\/}'s steam pipe. 
The assumption, in consequence, that the law is closely related to such 
engines, and to the circular processes involved, is of course not wrong, but 
it underestimates 
a  fundamental {\em logical-combinatorial-informational fact\/};
perhaps steam engines are to the second law what {\em telescopes\/} are for 
Jupiter's moons.

Carnot argued that the maximal efficiency of  a heat engine between 
two heat baths depended only on the two temperatures involved. 
(The derived formula motivated  Lord Kelvin to define the 
absolute temperature scale.)

{\em Rudolf Clausius'\/} (1822--1888)~\cite{clausius} version of the second law reads: {\em ``Es kann nie W\"arme aus einem k\"alteren in einen
w\"armeren 
K\"orper \"ubergehen, ohne dass eine andere damit zusammenh\"angende 
\"Anderung eintritt.''}~| ``No process can transport
heat 
from cold to hot and do no further change.''

{\em Lord Kelvin\/} (1824--1907)~\cite{kelvin}
formulated his own version of the second law and concluded~| in just
the next sentence~| that the
law may have consequences deeper than what was obvious at first sight:
{\em ``Restoration of mechanical energy without dissipation
[...]\ is impossible.
Within a finite period of time past, the earth must have been, within
a finite time, the earth must again be unfit for the habitation of man.''}

Also for Clausius, it was only a single thinking step
from his version of the law to concluding that 
all temperature differences in the entire universe will vanish
(the {\em W\"armetod\/}) and
that then, no
change will be possible anymore. 
He speaks of {\em a general tendency of nature for change into
  a specific direction:\/} ``Wendet man
dieses
auf das Weltall im Ganzen an, so gelangt man zu einer eigent\"umlichen
Schlussfolgerung, auf welche zuerst W.~Thomson [Lord Kelvin] aufmerksam
machte,
{\em nachdem er sich meiner Auffassung des zweiten Hauptsatzes
angeschlossen hatte}. Wenn [...]\ im Weltall die W\"arme stets das
Bestreben zeigt, [...]\ dass [...]\ Temperaturdifferenzen
ausgeglichen werden, so muss es sich mehr und mehr dem Zustand
ann\"ahern, wo [...]\ keine Temperaturdifferenzen mehr 
existieren.''~| in short: ``He was right after he had  realized that I had been
right: At the end, no temperature differences will be left in the universe.'' 

{\em Ludwig Boltzmann\/} (1844--1906) brought our understanding of the second 
law closer to combinatorics and probability theory (in particular, the
law of large numbers). His version is based on the fact 
that it is more likely to end up in a large set (of possible states)
than in a small one: The more
``microstates'' belong to a given ``macrostate,''
the more likely is it that you will find yourself in that macrostate. In other 
words, if you observe the time evolution of a system (by some reason 
starting in a very small, ``unlikely'' macrostate), then the ``entropy'' of
the system~| here simply (the logarithm of) the number of
corresponding microstates~| does not decrease.\footnote{Boltzmann
  imagined further
that the universe had started in a completely ``uniform'' state,
so the entire, rich reality perceived would  be a simple
fluctuation. (Note that the fact that this fluctuation is extremely
  unlikely is irrelevant if we can {\em condition on our existence},
    given our discussing this.)
He may have been  aware that this way of thinking leads straight
into {\em solipsism:\/}  ``My existence alone, simply {\em
  imagining\/} my environment, seems much more likely than the actual
existence of all people around me, let alone all the visible galaxies,
{\em etc}.''~| he killed himself in a hotel room in Duino, Italy; it
has been told that this was also related to ``mobbing'' by Mach in
Vienna. In any case,  we choose to comfort us today
with the somewhat religious
assumption that the universe initiated in a low-entropy state, called the
{\em big bang}.} 

The notion of macrostate and its
 entropy have been 
much debated.
Von Neumann remarked~\cite{uffink}:
``No one knows what entropy really is, so in a debate you will always
have the advantage.''
We 
aim at a version of the second law
avoiding this advantage:  a view without
probabilities or ensembles, but based on intrinsic, one-shot 
complexity instead. Crucial steps in that direction were made by {\em
  Zurek}~\cite{zurek}. 
We take a  Church-Turing view and follow  {\em Landauer\/}~\cite{landauer61} whose
role or, more specifically, whose choice of  viewpoint around the second 
law can be compared with {\em Ernst Specker\/}'s~\cite{specker} take on 
quantum theory: {\em logical.}

\subsection{Reversibility}

Landauer investigated the thermodynamic price of logical 
operations. He was correcting a belief by {\em John von Neumann\/}
that 
every bit operation required free energy $kT\ln 2$ (where $k$ is
Boltzmann's 
constant, $T$ the environmental temperature, and $\ln2$ owed to 
the fact that $2$ is not a natural but a logical constant).
According to Landauer~| and affirmed by {\em Fredkin and Toffoli's\/}
``ballistic computer''~\cite{ballistic}~|,
this limitation or condition only concerns (bit) operations which are 
logically {\em irreversible}, such as the AND or the OR. On
the positive side, it has been observed that
every function, bijective or not, can in principle be evaluated in a logically {\em reversible way,
  using
only ``Toffoli gates,'' i.e., made-reversible and then-universal AND gates\/};  its
computation can  be thermodynamically neutral:
It does not have to dissipate heat.

{\em Landauer's principle\/} states erasing (setting the
corresponding  memory cells to $0$)  $N$ bits costs 
$kTN\ln2$ free energy which must be dissipated as heat to the 
environment (of temperature $T$). This {\em dissipation\/} is crucial in the argument:
Heating up the environment compensates
for the {\em entropy loss \/} within the memory cell, realized as 
a physical system (spin, gas molecule, {\em etc.}). 

Let us consider the inverse process:
{\em Work extraction}. 
Bennett~\cite{bennetttoc} made the key contribution to 
the resolution of the paradox of {\em Maxwell's demon}. That demon 
had been thought of as violating the second law by adaptively 
handling a frictionless door with the goal of ``sorting a gas'' in
a container. Bennett took 
 the demon's memory (imagined to be in the all-$0$-state before sorting)
into account, which is in the end filled with
``random'' information, an expression of  the original state of the
gas. The 
growth of disorder {\em inside\/} the demon compensates for the order 
she creates {\em outside} ({\em i.e.}, in the gas)~| the second  law is
saved. The initial $0$-string is the demon's resource
allowing for her order creation.

If we break Bennett's argument apart in the middle, we  end up with
the {\em converse\/} of Landauer's principle: 
 The 
all-$0$-string has work value, {\em i.e.}, if we accept 
the price of the respective memory cells to become ``randomized'' in
the process,
we can  extract $kTN\ln2$ free energy from 
the environment (a~heat bath of temperature $T$). 
In a {\em constructivist\/} manner,
we choose to view the work-extraction process as an algorithm
which, according to the {\em Church-Turing hypothesis,} we
imagine as being carried out by a universal {\em Turing machine}. 
It then follows that the {\em work value of a string $S$\/} is closely related
to the possibility of lossless compression of that string: For any 
concrete data-compression algorithm, we can extract $kT\ln2$ times the length of $S$
(uncompressed) minus the length of its compression: {\em Work value is
 redundancy (in representation) of information.}
On the other end of the scale, the upper bound on work extraction
is linked to the ultimate compression limit: {\em Kolmogorov
  complexity}, {\em i.e.}, the length of the shortest program for the 
extraction demon (Turing machine) generating the string in question. This
holds  because a computation is logically reversible only
if it can be carried out in the other direction, step by step.

There is a direct connection between the work value and the {\em erasure 
cost\/} (in the sense of Landauer's principle) of a string. We assume here that for both processes, the
extraction demon has access to an additional string $X$ (modeling
prior ``knowledge'' about $S$) which serves as a catalyst and is to be 
unchanged at at the end of the process. 
For a string $S\in\{0,1\}^N$, let WV$(S|X)$ and EC$(S|X)$ be its work
value and erasure costs, respectively, given $X$. Then\footnote{Let
  $kT\ln 2=1$.}
\[
{\rm WV}(S|X)+ {\rm EC}(S|X)=N\ .
\]

To see this, consider first the combination extract-then-erase. Since 
this is {\em one specific way\/} of erasing, we have 
\[
{\rm EC}(S|X)\leq N- {\rm WV}(S|X)\ .
\]
If, on the other hand, we consider the combination erase-then-extract,
this leads to 
\[
{\rm WV}(S|X)\geq N- {\rm EC}(S|X)\ .
\]

Given the results on the work value discussed above, as well as this connection 
between the work value and erasure cost, we obtain the following
bounds
on the thermodynamic cost of erasing a string $S$ by a demon, 
modeled as a universal Turing machine ${\cal U}$ with initial 
tape content~$X$.
\\ \

\noindent
{\bf Landauer's principle, revisited.}
{\it 
Let $C$ be a computable compression function
\[
C\, :\, \{0,1\}^*\times \{0,1\}^* \longrightarrow \{0,1\}^*
\]
such that 
$
(A,B)\mapsto (C(A,B),B)
$
is injective. 
Then we have 
\[
K_{{\cal U}}(S|X)\leq {\rm EC}(S|X)\leq {\rm len}(C(S,X))\ .
\]}

Landauer's revised principle puts forward two ideas: First, the erasure
cost is an {\em  intrinsic, context-free, physical measure for 
  randomness\/}
(entirely independent of probabilities and counter-factual statements of
the form ``some value {\em could\/} just as well have been
{\em different},'' {\em i.e.}, removing one layer of speculation).
The idea that the erasure cost~| or the Kolmogorov complexity
related to it~| is a measure for randomness independent of
probabilities can be tested in a context in which randomness
has  been paramount: {\em Bell correlations\/}~\cite{bell} predicted
by quantum theory, see Section for details~\ref{bell}.

The second idea starts from the observation that
the  price for the {\em logical\/} irreversibility of the
erasure transformation comes in the form of 
 a {\em thermodynamic\/} effort.\footnote{Since the amount of the required free
  energy (and heat dissipation) is proportional  to the length of the
  best  compression of the string, the latter can be seen as a {\em
    quantification\/}
of the erasure transformation's irreversibility.}\
In an attempt to harmonize this somewhat {\em hybrid\/} picture, 
we invoke
Wheeler's~\cite{wheeler}
{\em ``It from Bit\/}: Every  {\em it\/}~| every particle, every field of force, even
  the space-time continuum itself~| 
derives its function, its meaning, its very existence entirely [...]\ from the apparatus-elicited 
answers to yes-or-no questions, binary choices, {\em bits}.''
This is an anti-thesis to Landauer's slogan, and we
propose the following synthesis of the two:  
If Wheeler suggests to look at the environment as being 
{\em information\/} as well, then Landauer's principle ends up to be
 read as: The necessary environmental compensation for 
the logical irreversibility of the erasure of $S$ is such that 
{\em the overall computation, including the environment, is
logically reversible: no information  ever gets completely lost.} 
\\ \ 

\noindent
{\bf Second law, Church-Turing view.}
{\it 
If reality is assumed to be computed by a Turing machine, then that
computation has the property of being injective: Nature computes with
Toffoli, but no AND or OR gates.
}
\\ \ \\
This fact is {\em a priori a\/}symmetric in time: The future
must 
uniquely determine the past, not necessarily {\em vice versa}. (This
is identical with {\em Grete Herrmann's\/}~\cite{gh} take on
causality.) In case the
condition
 holds also for the reverse time direction, the computation is {\em
  deterministic}, and {\em randomized\/} otherwise.

\subsection{Consequences}

If  logical reversibility is a simple computational version of a discretized second law, 
does it have implications resembling the traditional
versions of the  law?

\ \\

\noindent
{\bf Logical reversibility implies quasi-monotonicity.}
\\ \

\noindent
First of all, we find a ``Boltzmann-like'' form, {\em i.e.}, the
existence of a  quantity essentially monotonic in time. More
specifically, 
the logical reversibility of 
time evolution implies that the Kolmogorov complexity of the global state
at time $t$ can be smaller than the one at time $0$ 
only by at most $K(C_t)+O(1)$ if $C_t$ is a string encoding 
the time span $t$. The reason is that one possibility of describing 
the state at time~$0$ is to give the state at time~$t$, plus~$t$
itself; the rest is exhaustive search using only a constant-length
program
simulating forward time evolution (including possible randomness).

\ \\

\noindent
{\bf Logical reversibility implies Clausius-like law.}
\\ \

\noindent
Similarly, logical reversibility also implies
statements resembling the version of the second law due to {\em
  Clausius\/}: ``Heat does not spontaneously flow from cold to hot.''
The rationale here is that if we have a computation ~| the time
evolution~| using only (logically
reversible) Toffoli gates, then it is
{\em impossible\/} that this circuit computes  a transformation 
mapping a pair of strings to another pair  such that
the 
Hamming-heavier of the two becomes even heavier whilst the lighter
gets lighter. A~function {\em accentuating\/}  imbalance, instead of
lessening
it,
is not reversible, as a basic counting
argument shows. 
\\ \

\noindent
{\it Example.}
Let a circuit consisting of only Toffoli gates map an $N(=2n)$-bit string to 
another. We consider the map separately on the first and second 
halves and  assume the computed function to be 
conservative, {\em i.e.}, to leave the Hamming weight of the full
string
unchanged at~$n$ (conservativity can be seen as some kind of {\em first\/} law, {\em
  i.e.}, the preservation
of a quantity).
We  look at the excess of~$1$'s in one of the halves (which 
equals the deficit of $1$'s in the other). We observe that the
probability
(with respect to the uniform distribution over all strings of some 
Hamming-weight couple $[wn,(1-w)n]$) of the {\em imbalance
substantially growing\/} is exponentially weak. The key ingredient for the argument 
is the function's  injectivity.
Explicitly, the probability that the weight couple changes from
$\mbox{$[wn,(1-w)n]$}$
to $[(w+\Delta)n,(1-w-\Delta)n]$~| or more extremely~|, for $1/2\leq w<1$ and $0<\Delta\leq
1-w$, is
\[
\frac{{n \choose (w+\Delta)n}{n \choose (1-w-\Delta)n}}
{{n \choose wn}{n \choose (1-w)n}}
=2^{-\Theta(n)}\ .
\]
Note here that we even get the correct, exponentially weak ``error probability'' with
which the traditional second law can be ``violated.''

\ \\

\noindent
{\bf Logical reversibility implies Kelvin-like law.}
\\ \

\noindent
``A single heat bath alone has no work value.''
This, again, follows from a simple counting argument.
There exists no reversible circuit that, for general 
input environments (with a fixed weight~| intuitively: {\em heat
  energy\/}),
extracts redundancy, {\em i.e.}, work value,  and concentrates it in some pre-chosen
bit positions:  {\em Concentrated\/} redundancy is  {\em more\/}
  of it.
\\ \

\noindent
{\it Example.}
The probability that a fixed circuit maps a ``Hamming bath'' of
length~$N$
 and Hamming weight~$w$ to another such that the first $n$
positions 
contain all $1$'s and such that the Hamming weight of the remaining 
$N-n$ positions is $w-n$ (again, we are assuming conservation  here)
is 
\[
\frac{
{N-n \choose w-n}}{{N \choose w}}=2^{-\Theta(n)}\ .
\]

\subsection{Discussion and Questions}

We propose  a logical view of the second law of 
thermodynamics: {\em the injectivity or logical reversibility of time evolution}.
This is somewhat ironic as the second law has often been related  
to its exact opposite: {\em irreversibility}.\footnote{Since new
  randomness cannot be gotten rid of later, the equation reads:
``Logical reversibility plus randomness equals thermodynamic
{\em ir\/}reversibility.'' If you {\em can\/} go back logically
in a random universe, 
 then you certainly {\em
  cannot\/} thermodynamically.}
 It implies, within the 
Church-Turing view, $\mbox{Clausius-,}$ Kelvin-, and Boltzmann-like 
statements. 
We arrive at seeing a  law   {\em combinatorial in
nature\/}~| and its discovery in the context of  steam pipes  as 
 a historical incident. 

A logically reversible computation can still split up
paths~\cite{everett}.\footnote{Note that there is no (objective)
  splitting up, or randomness,
if time evolutions are unitary, {\em e.g.}, come from 
Schr\"odinger,  heat-propagation, or Maxwell's 
equations.
What is then the origin of the arrow of time?
The quantum-physical
version of injectivity is {\em Hugh Everett III's relative-state
  interpretation}. How do
we imagine the 
bridge from global unitarity to the subjective perception of
time 
asymmetry? 
When we looked above, with Landauer, at a closed {\em classical\/} system of two parts, 
then the (possible) complexity
deficit in one of them must simply be compensated in a corresponding
increase in the other. In Everett's view, this means that there can be 
low-entropy {\em branches of the wave function\/} (intuitively, yet too
na\"{\i}vely, called: parallel universes) as long as they are compensated by 
other, highly complex ones.}
This 
``randomness'' may bring 
in {\em objective\/} time asymmetry. What is then the exact mechanism by which randomness
implies that a {\em record\/} tells more about the past than about the
future? (Does it?)

\section{Bell Correlations and the
  Church-Turing Hypothesis}\label{bell}

We  test the obtained intrinsic notion of randomness,
in the form of erasure cost or Kolmogorov complexity,
with a physical phenomenon that we have already  mentioned above
as challenging {\em a-priori\/} causality: 
{\em ``non-local'' correlations\/} from quantum theory. In fact, 
randomness has  been considered crucial in the argument. We 
put this belief into question in its exclusiveness; at the same 
time we avoid in our reasoning  connecting  results of 
different measurements that, in fact, exclude each other (in other
words, we refrain from assuming 
so-called {\em counter-factual definiteness}, {\em i.e.}, that all these
measurement 
outcomes even {\em exist\/} altogether).\footnote{The {\em counter-factual\/} nature
of the reasoning
claiming ``non-classicality'' of quantum theory,
that was the main motivation in~\cite{pra}, has already been pointed out by
Specker~\cite{specker}: ``In einem gewissen Sinne geh\"oren aber auch
die scholastischen Spekulationen \"uber die {\em Infuturabilien\/} hieher,
das heisst die Frage, 
ob sich die g\"ottliche Allwissenheit auch auf Ereignisse erstrecke, 
die eingetreten w\"aren, falls etwas geschehen w\"are, was nicht geschehen
ist.''~---~``In some sense, this is also related to the scholastic
speculations on the {\em infuturabili}, {\em i.e.}, the question
whether divine omniscience even extends to what would have happened
if something had happened that did not happen.''}
For the sake of comparison, we first review the common, probabilistic,
counter-factual reasoning.

\subsection{Bell Non-Local Correlations}

Non-locality, manifested in violations of  {\em Bell
  inequalities}, expresses the impossibility to prepare parts of 
an entangled system simultaneously for {\em all possible measurements}. 
We look at  an idealized non-local correlation, the 
{\em Popescu-Rohrlich (PR) box\/}~\cite{pr}. Let $A$ and $B$ be the 
respective input bits to the box and~$X$ and~$Y$ the output bits; the
(classical)
bits satisfy
\begin{equation}
\label{piar}
X\oplus Y=A\cdot B\, .
\end{equation}
According to a result by Fine~\cite{fine}, the non-locality of the
system ({\em i.e.}, conditional distribution) $P_{XY|AB}$, which means 
that it cannot be written as a convex combination of products
$P_{X|A}\cdot P_{Y|B}$, is equivalent to the fact that there exists no
preparation for all alternative measurement outcomes $P'_{X_0X_1Y_0Y_1}$ such that 
\[
P'_{X_iY_j} =P_{XY|A=i,B=j}
\]
for all $(i,j)\in
\{0,1\}^2$. 
In this view, non-locality means that the outputs
cannot {\em exist\/}\footnote{What does it mean that a classical bit
  {\em exists\/}? Note first that
{\em classicality\/} of information implies that it can be measured without
disturbance and that
the outcome of a ``measurement'' is always the same; 
this makes it clear that it
 is an {\em idealized\/} notion requiring the classical bit to be
represented in a redundant  way over an {\em infinite\/} number of
degrees of freedom, as a thermodynamic limit. It makes thus sense 
to say that a  {\em classical bit $U$ exists\/}, {\em i.e.},
has taken  a
definite value.} before the inputs do. Let us make this
qualitative statement more precise. 
We assume a perfect PR box, {\em i.e.}, a system always satisfying 
$X\oplus Y=A\cdot B$. Note that this equation alone does not uniquely
determine $P_{XY|AB}$ since the marginal of~$X$, for instance, is not
determined. If, however, we additionally require {\em no-signaling}, 
then the marginals, such as $P_{X|A=0}$ or $P_{Y|B=0}$, must be perfectly 
unbiased under the assumption that all four $(X,Y)$-combinations, {\em
  i.e.},
$(0,0),(0,1),(1,0)$, and $(1,1)$, 
are possible. 
To see this, assume on the contrary that  $P_{X|A=0,B=0}(0)>1/2$. By the PR condition~(\ref{piar}),
we can conclude the same for $Y$: $P_{Y|A=0,B=0}(0)>1/2$. By
no-signaling,
we also have $P_{X|A=0,B=1}(0)>1/2$. Using symmetry, and no-signaling
again, 
we obtain both $P_{X|A=1,B=1}(0)>1/2$ and $P_{Y|A=1,B=1}(0)>1/2$.
This  contradicts the PR condition~(\ref{piar}) since {\em two bits which are  both
biased towards\/~$0$ cannot  differ with
certainty}. Therefore, our original assumption was wrong: The outputs 
{\em must\/} be perfectly unbiased. Altogether, this means that~$X$ as well as
$Y$ cannot exist ({\em i.e.}, take a definite value~--- actually,
there cannot 
even exist a classical value arbitrarily weakly correlated with one of
them) 
{\em before\/} the classical bit $f(A,B)$ exists
for some nontrivial deterministic
function $f\/ :\/ \{0,1\}^2\rightarrow \{0,1\}$. 
The paradoxical aspect of non-locality~--- at least if a causal
structure is  in   place~--- now consists of the fact
that {\em fresh} pieces of information {\em come to existence\/} in a
{\em spacelike-separated\/} manner  that are nonetheless {\em perfectly correlated}.

\subsection{Kolmogorov Complexity}

We  introduce the basic notions required for our alternative,
complexity-based view. 
Let ${\cal U}$ be a fixed universal Turing machine (TM).\footnote{The introduced asymptotic notions
are independent of this choice.}
For a finite
or infinite string $s$,  the {\em Kolmogorov
  complexity\/}~\cite{kol},~\cite{text} \mbox{$K(s)=K_{\cal U}(s)$} is the length
of the shortest program for~${\cal U}$ such that the machine outputs~$s$. Note that $K(s)$ can be infinite if~$s$ is.

Let $a=(a_1,a_2,\ldots)$
be an infinite string. Then
\[
a_{[n]}:=(a_1,\ldots,a_n,0,\ldots)\ .
\]
We study the asymptotic behavior of $K(a_{[n]})\, :\, {\bf
  N}\rightarrow {\bf N}$. For this function, we simply write~$K(a)$,
similarly $K(a\, |\, b)$ for $K(a_{[n]}\, |\, b_{[n]})$,  the latter being the length of 
the shortest program outputting~$a_{[n]}$ upon input~$b_{[n]}$. 
We write
\[
K(a)\approx n\ :\Longleftrightarrow\ \lim_{n\rightarrow\infty}
\frac{K(a_{[n]})}{n} =1\ .
\]
We call a string $a$ with this property {\em  incompressible}. 
We also use $K(a_{[n]})=\Theta(n)$, as well as
\[
K(a)\approx 0 :\Longleftrightarrow \lim_{n\rightarrow\infty}
 \frac{K(a_{[n]})}{n} = 0 \Longleftrightarrow K(a_{[n]})=o(n)\ .
\]
Note that {\em computable\/} strings $a$ satisfy $K(a)\approx 0$, and
that incompressibility is, in this sense, the extreme case of uncomputability.

Generally, for  functions $f(n)$ and $g(n)\not\approx0$, we write 
$f\approx g$ if $f/g\rightarrow 1$.
{\em Independence of $a$ and $b$\/} is  then\footnote{This is inspired
  by~\cite{cbc} (see also~\cite{google}), where (joint) Kolmogorov complexity~--- 
or, in practice, any efficient compression method~--- is used to
define 
a {\em distance measure\/} on sets of  bit strings (such as literary texts of
genetic information of living beings). The resulting structure in that
case is a distance measure, and ultimately a clustering as a binary tree.}
\[
K(a\, |\, b)\approx K(a)
\]
or, equivalently, 
\[
K(a,b)\approx K(a)+K(b)\ .
\]
If we introduce
\[
I_K(x;y):=K(x)-K(x\, |\, y)\approx K(y)-K(y\, |\, x)\ ,
\]
independence of $a$ and $b$ is $I_K(a,b)\approx 0$.

In the same spirit, we can define {\em conditional independence\/}: We
say that
{\em $a$ and $b$ are independent given $c$\/} if 
\[
K(a,b\, |\, c)\approx K(a\, |\, c)+K(b\, |\, c)
\]
or, equivalently, 
\[
K(a\, |\, b,c)\approx K(a\, |\, c)\ ,
\]
or
\[
I_K(a;b\, |\, c):=K(a\, |\, c)-K(a\, |\, b,c)\approx 0\ .
\]

\subsection{Correlations and Computability}

We are now ready to discuss non-local correlations with our context-free
randomness measure. The mechanism we discover is very similar 
to what holds probabilistically: If the choices of the measurements
are 
random (uncomputable) and non-signaling holds, then the outputs must
be random (uncomputable) as
well. We prove the following statement.
\\ \

\noindent
{\bf Uncomputability from Correlations.}
{\em There exist bipartite quantum states with a
behavior under measurements  such that if the sequences of setting
encodings are
maximally uncomputable
(incompressible), then the sequences of measurement results are 
uncomputable as well, even given the respective setting sequences.}
\\ \ 

\noindent
{\it Proof.}
We proceed step by step, starting with the idealized system of the PR~box. 
Let first $(a,b,x,y)$ be infinite  binary strings with
\begin{equation}
\label{piri}
x_i\oplus y_i= a_i\cdot b_i\ .
\end{equation}
Obviously, the intuition is that the strings stand for the inputs and
outputs of a PR box. Yet,  no dynamic meaning is attached
to the strings anymore (or to the ``box,'' for that matter) since
there is no
{\em ``free choice''\/} of an input and no generation of an output in function of the
  input; all we have
is a quadruple of  strings satisfying the PR condition~(\ref{piri}).
However, nothing prevents us from defining this
(static) situation to be {\em no-signaling\/}: 
\begin{equation}
\label{ns}
K(x\, |\, a)\approx K(x\, |\, ab)\mbox{\ \ \ and\ \  \ }K(y\, |\, b)\approx K(y\, |\, ab)\ .
\end{equation}

We argue that if the inputs
are  incompressible and independent, and no-signaling holds,
then the outputs must be uncomputable:
To see this,
assume now that $(a,b,x,y)\in (\{0,1\}^{\bf N})^4$ with $x\oplus y=a\cdot b$
(bit-wisely), no-signaling~(\ref{ns}), and 
\[
K(a,b)\approx 2n\ ,
\]
{\em i.e.}, the ``input'' pair is  incompressible. We  conclude 
\[
K(a\cdot b\, |\, b)\approx n/2\ .
\]
Note first  that $b_i=0$ implies $a_i\cdot b_i=0$, and second that 
any further compression of $a\cdot b$, given $b$, would lead to
``structure in $(a,b)$,'' {\em i.e.},   a
possibility of describing (programming)
$a$ given $b$ in shorter than $n$ and, hence,
 $(a,b)$ in
shorter than $2n$.
Observe now
\[
K(x\, |\, b)+K(y\, |\, b)\geq K(a\cdot b\, |\, b)
\]
which implies
\begin{equation}
\label{b1}
K(y\, |\, b)\geq K(a\cdot b\, |\, b)-K(x\, |\, b)\geq n/2-K(x)\ .
\end{equation}
On the other hand, 
\begin{equation}
\label{b2}
K(y\, |\, a,b)\approx K(x\, |\, a,b)\leq K(x)\ .
\end{equation}
Now, no-signaling~(\ref{ns}) together with~(\ref{b1}) and~(\ref{b2})
implies
\[
n/2-K(x)\leq K(x)\ ,
\]
and 
\[
K(x)\geq n/4 =\Theta(n)\ :
\]
(This bound can be improved
by a more involved argument~\cite{charles}.)
The string $x$ must be uncomputable. 

{\em A priori}, it is not overly surprising to receive uncomputable
outputs
upon inputs having the same property. Thus, we now turn our attention
to the {\em conditional\/} output complexities {\em given\/} the inputs:  We
consider the quantities 
$K(x\, |\, a)$ and $K(y\, |\, b)$. Note
first
\[
K(x\, |\, a)\approx 0\Leftrightarrow K(x\, |\, ab)\approx  K(y\, |\,
ab)\approx 0 \Leftrightarrow K(y\, |\, b)\approx
0\ , 
\]
{\em i.e.}, the two expressions  vanish simultaneously. 
We show that, in fact, they  both  fail to be of order $o(n)$.
To see this, assume $K(x\, |\, a)\approx 0$ and $K(y\, |\, b)\approx 0$.
Hence, there exist programs $P_n$ and $Q_n$ (both of length $o(n)$) 
for functions $f_n$ and $g_n$ with
\begin{equation}
\label{fgnl}
f_n(a_n)\oplus g_n(b_n)=a_n\cdot b_n\ .
\end{equation}
For fixed (families of) functions $f_n$ and $g_n$, asymptotically 
 how many $(a_n,b_n)$ can at most exist that satisfy~(\ref{fgnl})? 
The question boils down to a {\em parallel-repetition\/} analysis of
the {\em PR game\/}: A result by Raz~\cite{raz} implies that the number 
is of order $(2-\Theta(1))^{2n}$. Therefore, the two programs $P_n$ and
$Q_n$ together with the index, of length
\[
(1-\Theta(1))2n\ ,
\] 
 of the correct pair $(a,b)$ within the list 
of length $(2-\Theta(1))^{2n}$ lead to a program, generating $(a,b)$,
that has 
length
\[
o(n)+(1-\Theta(1))2n\ ,
\] 
in contradiction to the assumption of  incompressibility of $(a,b)$.

Unfortunately, 
perfect PR boxes are not predicted by
quantum theory. 
We show that correlations which {\em are\/} achievable in the laboratory~\cite{tit}
allow for the argument to go through; they are based on the {\em chained
  Bell inequality\/}~\cite{kent} instead of perfect PR-type non-locality.

To the chained Bell inequality belongs the following idealized system: 
Let $A,B\in\{1,\ldots,m\}$ be the inputs. We assume  the
``promise''
that $B$ is congruent to $A$ or to $A+1$ modulo $m$. Given this 
promise, the outputs $X,Y\in\{0,1\}$ must satisfy
\begin{equation}
\label{chain}
X\oplus Y={\chi}_{A=m,B=1}\ ,
\end{equation}
where ${\chi}_{A=m,B=1}$ is the characteristic function of the event
$\{A=m,B=1\}$. 

Barrett, Hardy, and Kent~\cite{kent} showed
that if $A$ and $B$ are random, then $X$ and $Y$ must be perfectly 
unbiased if the system is no-signaling. More precisely, they were 
even able to show such a statement from the gap between the 
error probabilities of the best classical~--- $\Theta(1/m)$~---
and quantum~--- $\Theta(1/m^2)$~--- strategies for winning this
game.

We assume 
$(a,b,x,y)\in (\{1,\ldots,m\}^n)^2\times (\{0,1\}^n)^2$ to be such
that the promise holds, and such that
\begin{equation}
\label{max}
K(a,b)\approx (\log m+1)\cdot n\ ,
\end{equation}
{\em i.e.}, the string $a||b$ is maximally incompressible 
 given the promise; the system is no-signaling~(\ref{ns});
the fraction of quadruples $(a_i,b_i,x_i,y_i)$, $i=1,\ldots,n$, satisfying~(\ref{chain}) 
is of order $(1-\Theta(1/m^2))n$. Then $K(x)=\Theta(n)$.

To see this, observe first that $K(a,b)$ being
maximal implies
\begin{equation}
\label{chib}
K(\chi_{a=m,b=1}\, |\, b)\approx \frac{n}{m}\ :
\end{equation}
The fractions of $1$'s in $b$ must, asymptotically,
be $1/m$ due to the string's incompressibility. If we condition on 
these positions, the string $\chi_{a=m,b=1}$ is
incompressible, since otherwise there would be the possibility of
compressing $(a,b)$.

Now, we have
\[
K(x\, |\, b)+K(y\, |\, b)+h(\Theta(1/m^2))n\geq K(\chi_{a=m,b=1}\, |\, b)
\]
since one possibility for ``generating'' the string $\chi_{a=m,b=1}$, 
from position $1$ to $n$, is to generate $x_{[n]}$ and $y_{[n]}$ as well as 
the string indicating the positions where~(\ref{chain})
is 
violated, 
the complexity of the latter being at most\footnote{Here, $h$
  is the binary entropy $h(x)=-p\log_2 p-(1-p)\log_2(1-p)$. Usually, $p$ is a probability, but $h$ is invoked here merely as an approximation for binomial coefficients.}\[
\log {n \choose \Theta(1/m^2)n}\approx h(\Theta(1/m^2))n\ .
\]

Let us compare this  with $1/m$: Although 
the binary entropy function has slope $\infty$ in 0, we have
\[
h(\Theta(1/m^2))<1/(3m)
\]
if $m$ is sufficiently large. To see this, observe first that the
dominant term of $h(x)$ for small $x$ is $-x\log x$, and second that
\[
c(1/m)\log(m^2/c)<1/3
\]
for $m$ sufficiently large.

Together with~(\ref{chib}), we now get
\begin{equation}
\label{c1}
K(y\, |\, b)\geq \frac{2n}{3m}-K(x)
\end{equation}
if $m$ is chosen  sufficiently large. On the other
hand, 
\begin{eqnarray}
K(y\, |\, ab)& \leq & K(x\, |\, ab)+h(\Theta(1/m^2))n\\
\label{c2}
&\leq & K(x)+\frac{n}{3m}\ .
\end{eqnarray}

Now,~(\ref{ns}), (\ref{c1}), and~(\ref{c2}) together imply 
\[
K(x)\leq \frac{n}{6m}=\Theta(n)\ ;
\]
in particular, $x$ must be uncomputable.  This concludes the proof. $\hfill\Box$

\subsection{Kolmogorov Amplification and the All-or-Nothing Nature of
  the Church-Turing Hypothesis}

The shown result implies  that {\em if\/} the experimenters are given access
to an incompressible number (such as $\Omega$~\cite{chaitin}) for choosing 
their measurement bases, {\em then\/} the measured photon (in a least one of
the two labs) is forced
to generate an uncomputable number as well, even given the 
string determining its basis choices. 

This is a similar observation as in the probabilistic realm, 
where certain ``free-will theorems'' have been formulated
in the context. In fact,  stronger statements hold there,
since non-local
correlations allow for {\em randomness amplification as well as
  expansion\/}~(see, {\em e.g.},~\cite{colrenamp}):
The randomness generated by the photons as their measurement output 
qualitatively {\em and\/} quantitatively {\em exceeds\/} what is required for
the choices of the measurement settings. This also holds in our complexity-based model: 
Indeed, it has been shown in~\cite{charles} that functionalities such
as
{\em Kolmogorov-complexity amplification and expansion\/} are
possible
using Bell correlations. 
The consequence is that there is 
either no incompressibility or uncomputability  at all in the world, or it is full of
it.
\\ \ 

\noindent
{\bf All-or-Nothing Feature of the Church-Turing Hypothesis.}
{\it 
Either no  device exists in nature allowing for producing uncomputable sequences,
or even a single photon can do it.
}

\section{Concluding Remarks and Open Questions}
The antagonism between the pre-Socratic philosophers {\em Parmenides\/} and
{\em Heraclitus\/} is still vivid in today's thinking traditions: The
Parmenidean line puts {\em logic\/} is the basis of space-time and dynamics~| in the
end 
all of physics. It has inspired researchers such as Leibniz, Mach, or Wheeler.
Central here is a doubt about {\em a priori\/}
absolute space-time causality: Is it  possible that these concepts only
emerge at a higher level of complexity, 
along with macroscopic, classical information?

Fundamentally opposed is the 
Heraclitean style, seeing
{\em physics and its objects\/} at the center: 
space, time, causality, and dynamic change 
is what all rests upon, including  logic,
computation, or information. To this tradition belong Newton, most
physicists including Einstein, the logician Gonseth, certainly
Landauer. 

According to {\em Paul Feyerabend\/}~\cite{feyer}, a specific
tradition
comes with its own criteria for success {\em etc.}, and it
can be judged from the standpoint of another (with those other criteria). In this spirit, 
it has been the goal of our discourse to build bridges between
styles, and to use
their 
tension 
to serve us.
This allowed, for instance,  to get  more insight 
into the second law of thermodynamics or the  ``non-local''
correlations from quantum theory. The latter 
 challenge our
established views of space and time; they, actually, have us look  back to the debate between Newton
and Leibniz and to question the path most of  science decided to take, at that time.

For the sake of a
final thought,
assume {\em \`a la\/} Leibniz 
that space, time, and causality do not exist prior
to {\em classical information\/}~| which we understand as an idealized notion
of  {\em macroscopically
  and highly redundantly represented\/} information; an~ideal classical bit
can then be measured without disturbance, copied, and easily
recognized as being classical. In this view, classicality is a {\em
  thermodynamic\/} notion. Thus the key to the {\em quantum measurement
process},
and  the problems linked to it, may lie within thermodynamics. (Yet, even if this
is
successful:
How come we observe correlations of pieces of {\em classical\/} information
unexplainable by any reasonable {\em classical\/} mechanism? How can quantum
correlations  and thermodynamic classicality~| {\em Bell \&
  Boltzmann\/}~| be reconciled?)

\ \\

\noindent
{\bf Acknowledgments.}
This text is based on a presentation at
the {\em 14th Annual Conference on
 Theory and Applications of Models of Computation (TAMC~2017)\/}
at the Universit\"at Bern. I am grateful to Gerhard J\"ager and all
the organizers for 
kindly
inviting me to give  a talk. 

I  thank 
Mateus Ara\'ujo, Veronika Baumann, \"Amin Baumeler, Charles B\'edard,
Claus Beisbart, Gilles Brassard,
Harvey Brown, Caslav Brukner, Harry Buhrman, Matthias Christandl, Sandro Coretti, 
Fabio Costa, Bora Dakic, Fr\'ed\'eric Dupuis, 
Paul Erker, Adrien Feix, J\"urg Fr\"ohlich, Manuel  Gil, Nicolas Gisin, 
Esther H\"anggi, Arne Hansen, Marcus Huber, Lorenzo Maccone, 
Alberto Montina, Samuel Ranellucci, Paul Raymond-Robichaud, Louis
Salvail, L.~Benno Salwey,  Martin Sch\"ule, Andreas Winter, and Magdalena Zych
for inspiring discussions, and the  {\em Pl\"af\"a-Einstein\/} as well as the {\em
  Reitschule\/}  for their inspiring atmosphere.

	This research  is supported by the Swiss National Science
        Foundation (SNF), the National Centre of Competence in
        Research ``Quantum Science and Technology'' (QSIT), 
the COST action on Fundamental Problems in Quantum Physics, and by the
{\em Hasler Foundation}.

\end{document}